# Quantum-critical phase out of frustrated magnetism in a strongly correlated metal

Hengcan Zhao[1,2], Jiahao Zhang[1,2], Meng Lyu[1,2], Sebastian Bachus[3], Yoshifumi Tokiwa[3], Philipp Gegenwart[3], Shuai Zhang[1,2], Jinguang Cheng[1,2], Yi-feng Yang[1,2], Genfu Chen[1,2,4], Yosikazu Isikawa[5], Qimiao Si[6], Frank Steglich[1,7,8], and Peijie Sun[1,2,4]

Email, pjsun@iphy.ac.cn; steglich@cpfs.mpg.de; qmsi@rice.edu

*[1] Beijing National Laboratory for Condensed Matter Physics, Institute of Physics, Chinese Academy of Sciences, Beijing 100190, China*
*[2] School of Physical Sciences, University of Chinese Academy of Sciences, Beijing 100049, China*
*[3] Experimental Physics VI, Center for Electronic Correlations and Magnetism, Institute of Physics, University of Augsburg, 86135 Augsburg, Germany*
*[4] Songshan Lake Materials Laboratory, Dongguan, Guangdong 523808, China*
*[5] Graduate School of Science and Engineering, University of Toyama, Toyama 930-8555, Japan*
*[6] Department of Physics and Astronomy, Rice University, Houston, TX 77005, USA*
*[7] Max Planck Institute for Chemical Physics of Solids, 01187 Dresden, Germany*
*[8] Center for Correlated Matter, Zhejiang University, Hangzhou, Zhejiang 310058, China*

**Abstract:**

**Strange-metal phenomena often develop at the border of antiferromagnetic order in strongly correlated metals. It has been well established that they can originate from the fluctuations anchored by the point of continuous quantum phase transition out of the antiferromagnetic order, i.e., a quantum critical point. What has been unclear is how these phenomena can be associated with a potential new *phase* of matter at zero temperature. Here we show that magnetic frustration of the 4*f*-local moments in the distorted Kagome intermetallic compound CePdAl gives rise to such a paramagnetic quantum-critical phase. Moreover, we demonstrate that this phase turns into a Fermi liquid through a Mott-like crossover; in a two-dimensional parameter space of pressure and magnetic field, this crossover is linked to a line of zero-temperature 4*f*-electron localization-delocalization phase transitions at low and moderate pressures. Our discovery motivates a new design principle for strongly correlated metallic states with unconventional excitations that may underlie the development of such effects as high temperature superconductivity.**

Geometrical frustration in quantum-spin systems gives rise to quantum fluctuations which may suppress long-range magnetic order and cause a quantum-spin-liquid ground state [1]. This notion is traditionally associated with insulating magnets only. There has been increasing recognition, however, that geometrical frustration is also important to bad metals that host local moments, such as strongly correlated *f*-electron metals [2-7], which provide a prototype setting for magnetic quantum phase transitions (QPTs) [8-12]. From general theoretical considerations, the magnetic Ruderman, Kittel, Kasuya, Yoshida (RKKY) interactions on a geometrically frustrated lattice generate quantum fluctuations that interplay with those produced by the competing Kondo interactions [6-7,13]. The QPT can then involve not only antiferromagnetic (AF) order and a paramagnetic heavy Fermi liquid, but also a potential metallic spin-liquid phase. While the latter phase has been argued to exist in some Kondo systems [2-5,14,15], how it can develop from magnetic frustration in a Kondo lattice has not been demonstrated so far.

The distorted Kagome-lattice system CePdAl with local quantum spins ($S_{eff}$ = ½) in the Kelvin range is well suited for this study [16,17]. Due to geometrical frustration only 2/3 of these local spins are involved in the magnetic ordering, and the frustration-assisted short-range order sets in at $T_m(B)$, significantly above the ordering temperature $T_N(B)$ [17, 18], see Fig. 1a. In the field-induced paramagnetic state, well separated from a first-order AF transition at $B_{cp} \simeq$ 4.2 T, a continuous QPT (or quantum critical point, QCP) has been identified [18] at $B_0^* \simeq$ 4.6 T, cf. also Fig. S1 in Supplementary Information (SI). At this field, about 75% of the magnetization is already saturated [19]. At $T = 0$ and on increasing the field through $B_0^*$, a partial delocalization of the 4$f$ - states takes place, which corresponds to the $T \rightarrow 0$ extrapolation of a Mott-type (localization-delocalization) crossover line $B^*(T)$ [18]. In addition, $B_0^*$ is the $T \rightarrow 0$ terminating point of two other crossover lines, i.e., into a Fermi-liquid phase at $B > B_0^*$ and into a frustration-assisted metallic spin-liquid state at $B_{cp} < B < B_0^*$. In contrast to the behavior of other AF Kondo systems [14, 15], for CePdAl non-Fermi-liquid (nFL) phenomena reflecting quantum criticality could not be observed in the electrical resistivity in the paramagnetic low - $T$ phase near $B_0^*$. This was ascribed to the large Zeeman splitting of the crystal-field (CF) ground-state doublet of the Ce - 4$f$ shell which causes a correspondingly large reduction of spin-flip scattering and an enhanced temperature dependence of the resistivity [18].

To avoid the effect of Zeeman splitting, here we study CePdAl at hydrostatic pressure $p$ up to about 2 GPa, using a single crystal from the same batch as employed in ref. 18. In Ce-based heavy-fermion systems, pressure is well known to enhance the Kondo coupling between the 4$f$ and conduction electrons and to weaken the magnetic order. Therefore, the magnetic QPT can be studied by the pressure alone or in combination with a lower magnetic field and, thus, at a smaller ratio of the Zeeman splitting to the Kondo energy. As a result, the nFL phenomena related to quantum criticality may eventually become fully resolved, through measurements such as of the electrical resistivity as a function of temperature. Two immediate questions arise: (1) will one observe signatures for Mott-type quantum criticality even at elevated pressures and, if so, (2) to which extent is the latter intertwined with the magnetic frustration?

The $T - p$ phase diagram at zero magnetic field is shown in Fig. 1a. Hydrostatic pressure indeed suppresses the AF order in CePdAl at $p_{c1} \approx$ 0.9 GPa, as already reported by Goto et al [19]. In contrast to the typical observation of a funnel-shaped quantum critical regime centered at a QCP, we observe nFL behavior over a finite pressure *range* between $p = p_{c1}$ and $p_{c2} \simeq$ 1.7 GPa. Here, the temperature dependence of the electrical resistivity as a function of $T$ follows a power law, $\Delta\rho(T) = \rho(T) - \rho_0 = A'T^n$, with the exponent $n$ being smaller than the Fermi-liquid value $n = 2$, see Figs.1c inset and S2a, as well as the color code in Fig. 1a. The upper cutoff, $T_m(p)$, of this nFL regime is smoothly connected to the afore-mentioned crossover line $T_m(B)$ at lower pressure; measurements of magnetoresistivity [18] and entropy [17] at ambient pressure have implicated the latter to be the onset of frustration-assisted short-range order. The crossover scale $T_m(p)$ vanishes at $p = p_{c2}$. A Fermi-liquid-like temperature dependence of the resistivity, $\Delta\rho(T) = AT^2$, with large $A$-coefficient, is observed in the antiferromagnetically ordered regime $p < p_{c1}$, similar to what was found in the AF low-field phase of YbRh$_2$Si$_2$ [20], as well as at $p > p_{c2}$.

Additional evidence beyond resistivity for this quantum-critical phase is provided by the specific-heat of Ni-substituted CePdAl at ambient pressure. Fig. 1b shows the 4$f$-derived specific heat, presented as $C_{4f}/T$ vs $T$, for Ce(Pd$_{1-x}$Ni$_x$)Al single crystals, with $x$ varying between 0 and 0.4. Ni-substitution for Pd has been shown to act as chemical pressure [21], where a critical Ni concentration $x_{c1} = 0.144$ corresponds to $p_{c1}$ and, assuming a linear relationship, $x_{c2} \simeq 0.27$ is expected to correspond to $p_{c2}$. Upon low Ni substitution the magnetic phase-transition anomaly is rapidly broadened and shifted to lower temperatures [22]. For $x = 0.2$, which lies in between $x_{c1}$ and $x_{c2}$, a nFL–like, divergent $T$– dependence is seen in $C_{4f}(T)/T$, in line with the corresponding $\rho(T)$ curve that reveals a $T^{1.6}$ dependence below about 0.5 K (inset of Fig. 1b). For $x = 0.4$, which is much larger than $x_{c2}$, $C_{4f}(T)/T$ saturates below about 0.3 K, and a $T^2$ dependence of the resistivity is observed at $T < 0.8$ K (inset of Fig. 1b). The zero-temperature values of $C_{4f}(T)/T$, extrapolated from the measured low-$T$ data, vary between 0.25 and more than 1.5 J/K$^2$mole, illustrating heavy-fermion behavior.

To further characterize the nFL phase, we show in Fig. 1c for the pure CePdAl the pressure dependences of the coefficients $A$ and $A'$ of the power law dependences in $\Delta\rho(T) = AT^2$ and $A'T^n$ ($n < 2$) as well as of the residual resistivity $\rho_0$ at $B = 0$. All three quantities exhibit a sharp peak at $p_{c1}$, where the resistivity exponent becomes smallest ($n \simeq 1.4$); $n$ which is 2 in the AF phase, shows a relatively sharp decrease for $p \geq p_{c1}$ and reaches 2 again close to $p_{c2}$ (see inset of Fig. 1c). In contrast to $p_{c1}$, $p_{c2}$ can only be identified from a very weak anomaly, e.g., a small and broadened hump in $\rho_0(B)$.

Fig. 2 displays the $T$- dependence of the zero-field resistivity between 0.06 and 1.5 K for three different pressure values within the nFL phase, see also Fig. S2a (SI). These results reveal an increasing "quantum critical strength" on releasing the pressure inside this phase. Specifically, the apparent linear $T$ – dependence of $\rho(T)$ found at elevated temperatures turns into a power-law behavior, $\Delta\rho(T) \sim T^n$, $1 < n < 2$, in the low – $T$ regime: The crossover temperature decreases upon decreasing pressure, i.e., on the approach of the QCP at $p_{c1} \simeq 0.9$ GPa; at the same time, the exponent $n$ decreases from 1.7 (1.44 GPa) to 1.6 (1.15 GPa) and reaches the critical value $n = 1.4$ (0.93 GPa) slightly above $p_{c1}$. A corresponding trend is seen in the specific-heat results of the CePdAl samples with Ni substitution $x = 0.2$ and 0.14, respectively, presented in Fig. S3 (SI). Here, the coefficient $C_{4f}(T)/T$ follows a $-\ln T$ dependence at elevated temperatures and crosses over to a weaker increase at lower temperatures: Again, the crossover temperature decreases upon approaching the critical concentration $x = 0.144$. The existence of these singular transport and thermodynamic properties over an extended range of pressure provides an unprecedented level of evidence and characterization for a quantum critical phase. Compared to heavy fermion metals without apparent geometrical frustration, where this type of phase has been implicated [14,15,23], the frustrated geometry of the crystalline lattice in CePdAl apparently makes possible the striking evidence for a quantum-critical phase in the continuous evolution of low-temperature singularities of both transport and thermodynamic properties.

In addition to bringing about a quantum-critical phase, the geometrically frustrated nature of CePdAl provides an opportunity to elucidate its interplay with the Kondo coupling. To do so, it is necessary that the system is tuned by more than one control parameter. With the further motivation provided by the results of field tuning at ambient pressure [16-18], we proceed to study the combined effect of pressure and magnetic field $B$. Importantly, under pressure, the increased Kondo coupling will enhance the "Kondo broadening" (see supplementary Fig. S9 and related discussion) of the Zeeman states and diminish, in the measured resistivity vs temperature dependence, the effect of Zeeman splitting, thereby revealing the true nFL behavior. For the finite field $B = 2$ T, the $T$ - $p$ phase diagram is shown in Fig. 3a, which is qualitatively similar to that for $B = 0$ in Fig. 1a. Different to the nFL-type $T^n$ dependence of the resistivity with $1.4 < n < 2$ over a largely extended pressure range $0.9 – 1.7$ GPa (i.e., $p_{c1} – p_{c2}$) for $B = 0$, a low – $T$ power law with $n = (1 \mp 0.1)$ is observed for $B = 2$T within the range $0.8$ GPa $< p < 1.0$ GPa, a much reduced pressure window.

Complementarily, in Fig. 3b, the $T$-$B$ phase diagram for $p = 0.82$ GPa, slightly below $p_{c1}$, is displayed. When compared with the corresponding phase diagram at ambient pressure (see Fig. 2 in Ref. 18), the most striking difference turns out to be an exponent $n = (1 \mp 0.1)$ in the power – law temperature dependence of $\rho(T)$ at $B = 3$ T (see Figs. S6a and d) at which field this almost linear temperature dependence is observed from $0.4$ K all the way down to $T \simeq 0.06$ K, the lowest accessible temperature. Such a nearly $T$ - linear resistivity extends to about $B_0^* \approx 3.5$ T, see Fig. S6a. In contrast, no nFL behavior ($n < 2$) could be resolved at $p = 0$ [18], as mentioned before. As is clearly seen in Fig. 3b, at $B > B_0^*$ the apparent quadratic $T$ – dependence of $\rho(T)$ indicates a Fermi – liquid phase.

In the following, we will discuss results which highlight the existence of a $B_0^*(p)$ line, at which the 4$f$-electrons delocalize in the zero-temperature limit [24-28]. Fig. 3c shows the evolution of the $A$-coefficient of the $T^2$ dependence of the resistivity as a function of the magnetic field near and above the critical pressure $p_{c1}$. For $p = 0.82$ GPa, there is a strong increase from both the AF and paramagnetic side suggesting a potential divergence of $A(B)$ at a Kondo – destroying AF QCP, $B_0^*(p = 0.82$ GPa$) \simeq 3.5$ T. Note that for this pressure the resistivity exhibits an almost linear temperature dependence of $\rho(T)$ between $B = 3$ T and $3.5$ T to the lowest temperature as mentioned above (supplementary Fig. S6a) [29]. Likewise, a $T$ - linear $\rho(T)$ is observed at $B_0^*(p = 0.93$GPa$) \simeq 2$ T, too, see Fig. S6b and d (SI). The so-derived values of $B_0^*(p)$ reasonably follow the ones read off the $B_0^*(p)$ curve in the inset of Fig. S5a, as derived from the inflection points/maxima in $\chi'(B)$, measured at fixed pressure values and registered at $T = 0.08$ K (cf. arrows in Fig. S5a). At pressure values in excess of $0.93$ GPa, the resistivity shows nFL behavior at low fields, see Figs. 1a and S6c (SI), but follows a $T^2$ law at fields $B$ sufficiently larger than $B_0^*(p)$. In this latter range, one still observes an increase of $A(B)$ upon decreasing field, this increase becoming continuously weaker upon increasing pressure and extending to lower crossover fields $B_0^*(p)$, see Fig. 3c - again illustrating that, inside the quantum critical phase, the quantum critical singularity becomes more pronounced upon decreasing pressure.

We can now put together the combined pressure and field tunings into a zero-temperature phase diagram, Fig. 4a. Here, the color code illustrates the exponent of the low $T$ – power law dependence of $\rho(T)$ only at fields at which the Zeeman splitting can be neglected, see SI. Most strikingly, this ($T$ = 0) $p$ - $B$ phase diagram reveals a quantum critical phase between the Mott line $B_0^*(p)$ and the line of AF quantum phase transitions. The $B_0^*(p)$ line contains several segments: (a) from the ambient pressure to about $p$ = 0.3 GPa where, most likely, a line of continuous QPTs or QCPs exists (see SI); (b) from $p$ = 0.8 GPa to slightly below $p$ = 1 GPa, where the $T$-linear resistivity persists to the lowest temperatures (Fig. S6d) and the $A$ - coefficient of the $T^2$-resistivity component diverges as the line is approached from both sides (see Fig. 3c), both observations implicating the continuation of the line of QCPs from the low-pressure part; and, finally, (c) for pressures from $\simeq$ 1 GPa to the terminating pressure ($p_{c2} \simeq 1.7$ GPa) of the line, in which part it corresponds to a crossover that separates the non-Fermi liquid [$\Delta\rho(T) \sim T^n$, $n$ < 2] and Fermi liquid [$\Delta\rho(T) \sim T^2$] regimes, as shown by the color code in Fig. 4a, and where the $A$ - coefficient of the $T^2$ term in $\rho(T)$ is maximized but does not diverge (Fig. 3c). Across the Mott line, even though measurements of angle-resolved photoemission or quantum-oscillation are essentially impossible because of the low temperature and the finite but not sufficiently large magnetic fields involved, a jump in the Fermi surface across the segment (a) is evidenced by the combined magneto-transport and magnetic susceptibility measurements at ambient [18] as well as finite but relatively small pressure (SI). By continuity, we conjecture that this jump continues up to and across segment (b) of the Mott line. For the AF quantum phase transitions, at $p \simeq$ 0.8 GPa and $B \simeq$ 3T, one may likely observe in future studies an AF multi-critical point; for fields below this point, the AF transition is continuous, while for fields above it, the outermost transition is of first order [17,18]. Finally, we recall that it is intriguing to observe the magnetically ordered phases of CePdAl to show heavy Fermi-liquid behavior, similar to the case of YbRh$_2$Si$_2$ [20].

Our results can be viewed from the perspective of a ($T$ = 0) 'global' phase diagram [6-7, 30] (Fig. 4b), which is specified in terms of two quantities: the ratio of the Kondo interaction ($J_K$) to the RKKY interaction ($I$), and the degree of quantum fluctuations of the local-moment magnetism ($G$). The tuning parameters pressure and magnetic field don't affect the geometrical frustration of the (distorted) Kagome lattice which dictates a large value of $G$. However, they modify the magnetic frustration and further tune $G$ by changing the effective range of the RKKY interaction, while also varying the $J_K/I$ ratio. This means that the application of pressure and field, which are non-universal quantities, results in two different cuts within the orange part of the global phase diagram. We can then qualitatively project the tunings in the $p$ - $B$ plane of Fig. 4a onto the $G$ - $J_K/I$ plane of Fig. 4b and interpret the observed non-Fermi liquid phase in between the afore-mentioned cuts in terms of the frustration-induced paramagnetic phase with Kondo destruction [24-27, 31] and a small Fermi surface ($P_S$). This implies another significance of our results, namely that the observed transition from the magnetic order to the paramagnetic nFL phase represents a novel type of metallic AF QCP of the local moments; here, the local moments are Kondo-coupled to the conduction electrons at the Hamiltonian level but the ground state on either side of the transition is Kondo-destroyed. This type of quantum criticality, while inherent to the proposed global phase diagram, has proven difficult to be observed [32]. Whether measurements at even lower temperatures can isolate the singularities associated with this QCP from those of the nFL phase associated with Mott line, $B_0^*(p)$, is an intriguing open question for future studies.

To conclude, we have presented evidence for a quantum-critical phase in the geometrically frustrated heavy fermion metal CePdAl. This finding has been made possible by the extensive tuning of both pressure and magnetic field. We found this phase to be located between two lines of QCPs, one for AF order and the other one for $f$ - electronic delocalization-localization in the form of Kondo-destruction; for the latter, the line of continuous phase transitions connects with a line of crossovers upon increasing pressure and decreasing magnetic field. Our results demonstrate geometrical frustration as a means of creating strange metals [33], whose unusual electronic excitations underlie the exotic properties of quantum materials such as the high temperature cuprate superconductors [34].

**Methods**

Single crystals of CePdAl and its Ni doped homologues were prepared by the Czochralski method in an argon gas atmosphere of 5 Torr using a tungsten crucible in an induction furnace [16]. The samples were oriented by the Laue backscattering and/or x-ray diffraction method, and then cut into bar-shape by a precision diamond wire saw. For all the measurements involved in this work, the electrical current was applied perpendicular to the $c$ axis and magnetic field parallel to $c$ axis, which is the Ising-like, easy magnetic axis. In order to produce hydrostatic pressures, we used a self-clamped piston-cylinder pressure cell made of nonmagnetic CuBe and NiCrAl, with glycerol as the pressure-transmitting medium. The generated pressure was monitored by the superconducting transition temperature of a small piece of Sn mounted together with the sample inside the pressure cell. The pressure cell was further loaded into a low-temperature $^3$He-$^4$He dilution refrigerator for transport and magnetic measurements down to $T$ = 0.06 K. For both electrical resistivity and Hall measurements, a CePdAl sample of dimension 0.2 x 0.5 x 2 mm$^3$ was used; for magnetic susceptibility measurements, we employed a cube-like sample (about 1 mm$^3$) cut from the same batch. For specific heat measurements on Ni-doped samples, a heat relaxation method was employed in a dilution refrigerator [22].

**Acknowledgements**

This work was supported by the Ministry of Science and Technology of China (Grant Nos. 2015CB921303 & 2017YFA0303100), the National Natural Science Foundation of China (Grant Nos. 11774404 and 11474332), the Chinese Academy of Sciences through the Strategic Priority Research Program (Grant No. XDB07020200), and a fund from the Science and Technology on Surface Physics and Chemistry Laboratory (No. 01040117). Work at Augsburg was supported by the German Research Foundation (DFG) under the auspices of TRR 80, while work at Dresden was partly supported by the DFG Research Unit 960. The work at Rice was supported in part by the NSF Grant DMR-1611392 and the Robert A. Welch Foundation Grant C-1411. Q.S. acknowledges the hospitality and support by a Ulam Scholarship from the Center for Nonlinear Studies at Los Alamos National Laboratory, and the hospitality of the Aspen Center for Physics (NSF, PHY-1607611).

**Figures**

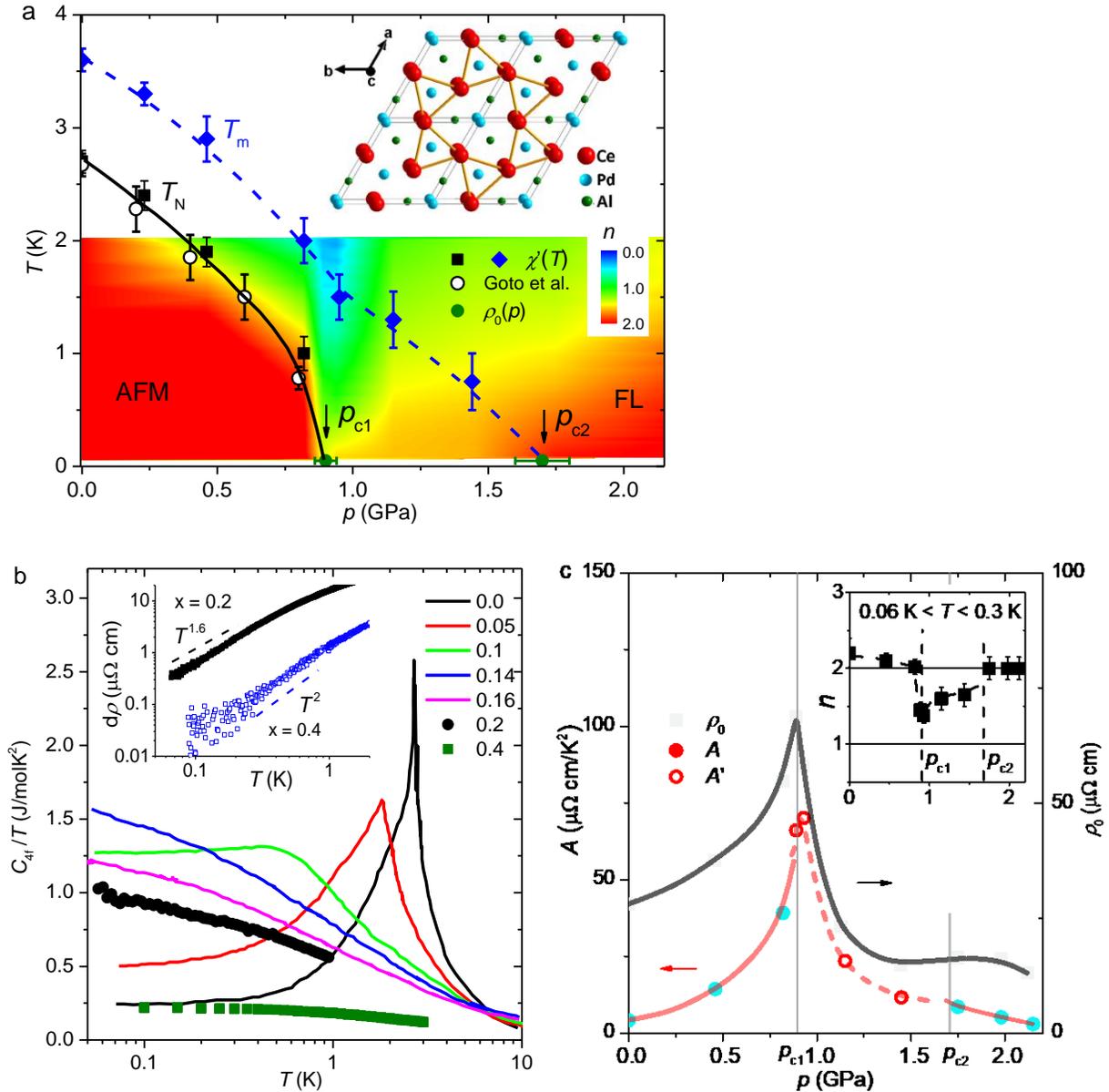

**Fig. 1. Results at zero field illustrating the emerging quantum critical phase.** (**a**) $T$-$p$ phase diagram for $B = 0$ on top of a color-coded map for the exponent $n$ of $\Delta\rho(T) \sim T^n$. The values of $T_N$ and $T_m$ are derived from the the inflection points and maxima, respectively, of the (real part of the) ac susceptibility $\chi'(T)$ curves, displayed in supplementary Fig. S2b. Open circles mark the Néel temperatures derived from the dc susceptibility measurements by Goto et al. [18]. The phase space at $T > 2$ K is left white because $\rho(T)$ was typically measured only up to 2 K in a dilution fridge. Inset: a sketch of the distorted Kagome lattice. (**b**) 4$f$-derived specific heat divided by $T$, $C_{4f}/T$, as a function of $T$, for CePd$_{1-x}$Ni$_x$Al with $x$ up to 0.4. The data for $x$ = 0.2 and 0.4, which are beyond the critical concentration $x \approx 0.144$ [19], were first determined in this work, while the other curves were reproduced from [21]. For $x$ = 0.2, a logarithmic increase of $C_{4f}(T)/T$

is clearly seen down to 0.06 K. By contrast, a flat and large $C_{4f}(T)/T$ below 0.5 K is found for $x$ = 0.4, indicative of heavy Fermi-liquid behavior as documented by the large size of $C_{4f}(T)/T$ at low temperatures. Inset shows $\Delta\rho$ vs $T$ in a double-log scale; here $\Delta\rho \propto T^{1.6}$ for $x$ = 0.2 and $\Delta\rho \propto T^2$ for $x$ = 0.4 are observed. (**c**) Coefficients $A$ and $A'$ in $\Delta\rho(T) \sim T^n$, with $n$ = 2 and $n < 2$, respectively, as well as the residual resistivity $\rho_0$ as a function of $p$. The large size of the $A$ coefficient documents a heavy Fermi-liquid phase. Inset shows $n$ vs $p$ estimated at the lowest temperatures (0.06 - 0.3 K). Note that $n \approx 1.4$ at $p_{c1}$, and that the FL behavior with $n$ = 2 is restored at $p > p_{c2}$.

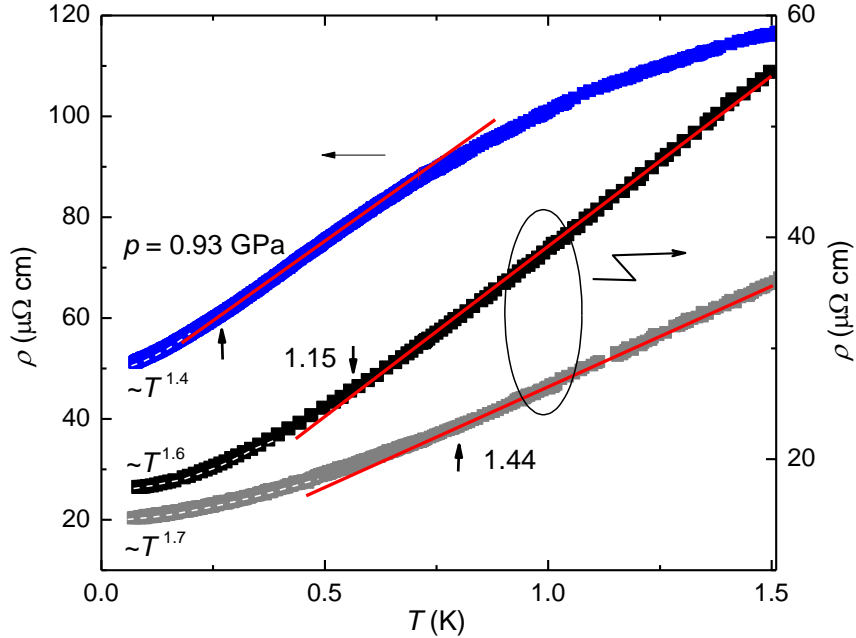

**Fig. 2. Pressure dependence of $\rho(T, B = 0)$ as a measure of the "quantum critical strength" within the quantum critical phase.** For all the three pressures ($p$ = 0.93, 1.15 and 1.44 GPa) within the quantum critical phase, a linear – in - $T$ resistivity is observed at elevated temperatures; below a crossover temperature (marked by arrow), $\Delta\rho(T)$ changes to $\sim T^n$ ($1 < n < 2$). Both the values of $n$ and crossover temperature decrease upon lowering the pressure, indicating an increasing "quantum critical strength". Note that for $p$ = 0.93 (close to $p_{c1}$) the intermediate temperature range, where $\Delta\rho(T) \sim T$, is substantially limited at the high – $T$ side by the combined action of Kondo-lattice coherence and significant AF short-range correlations which become visible already at $T > 1$ K (see SI).

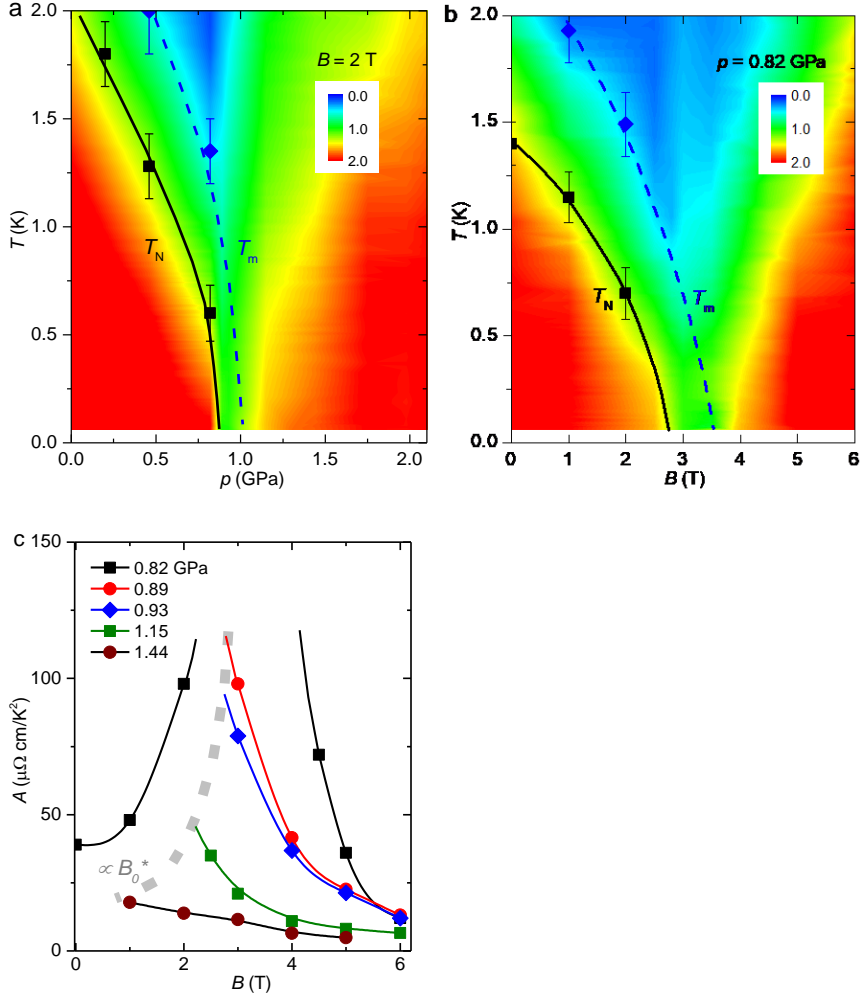

**Fig. 3. Non-Fermi-Liquid behavior studied at combined control parameters of magnetic field and hydrostatic pressure.** (**a**) $T$-$p$ phase diagram at finite field ($B = 2$ T). nFL behavior characterized by an almost linear-in-$T$ resistivity exists between the $T_N(p)$ and $T_m(p)$ lines, as evidenced by the green coding in the $n$ map. For the $\rho(T)$ curves measured at varying pressures and fields, we refer to Fig. S6. (**b**) $T$-$B$ phase diagram at finite pressure ($p = 0.82$ GPa). Similar to (**a**), here nFL is also observed between the $T_N(p)$ and $T_m(p)$ lines. This is to be compared with the complex $T$-$B$ phase diagram at ambient pressure, where multiple first-order metamagnetic transitions appear, but nFL is absent [17]. (**c**) The $A$ - coefficient estimated from the Fermi-liquid $T^2$-dependence of $\rho(T)$ as a function of $B$ for different pressures. For $p = 0.82$ GPa, an incipient divergence of $A(B)$ is observed from below and above $B_0^* \simeq 3.5$ T. For both $p = 0.89$ and $0.93$ GPa (close to $p_{c1}$), a divergence of $A(B)$ towards $B_0^* \simeq 2$ T from the paramagnetic side may still be anticipated. At higher pressure values, nFL behavior is found on the low-field side. However, $\rho(T)$ exhibits a $T^2$ dependence sufficiently above $B = B_0^*(p)$. Here, the field where $A(B, p = $ const.) takes the largest values decreases with increasing pressure and vanishes at $p \simeq p_{c2}$. The dashed gray line indicating the locations of these largest $A$ - coefficients is in good agreement with $B_0^*(p)$ as derived independently from other probes (see text).

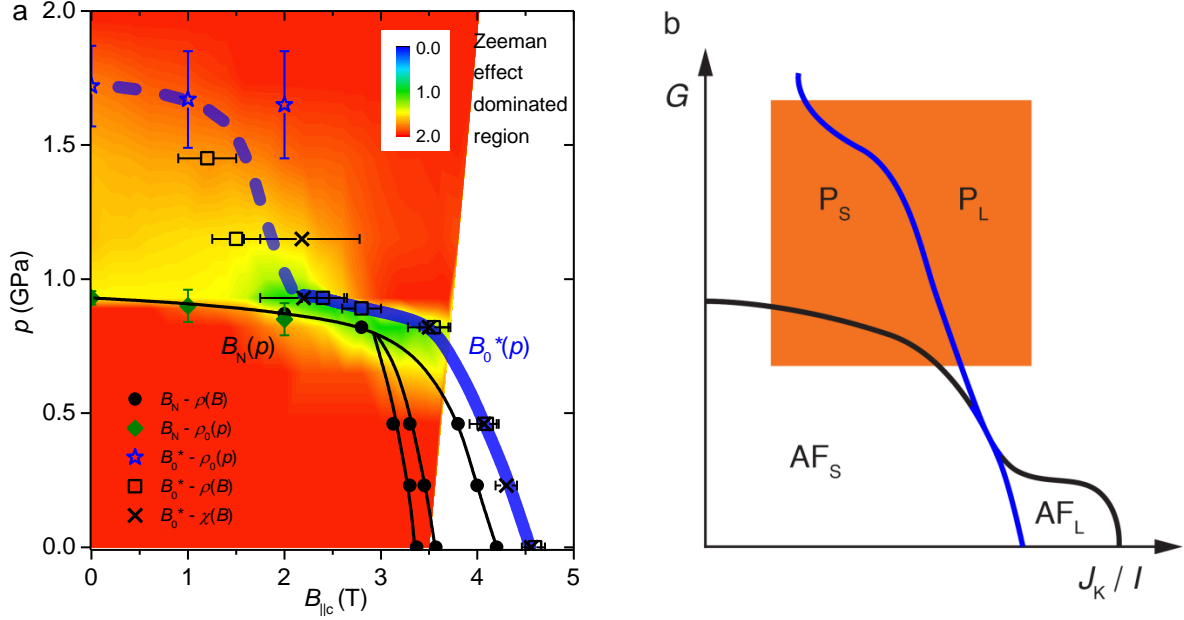

**Fig. 4. Experimental $p - B$ and schematic global $G - J_K/I$ phase diagrams at zero temperature. (a)** $p - B$ phase diagram for magnetically frustrated CePdAl at $T = 0$ with quantum phase transitions (solid lines) and crossover (dashed thick line) as extrapolated from finite temperature results. Color code for fields below $\simeq 3.5$ T ($p = 0$) and $\simeq 3.7$ T ($p = 0.82$ GPa), where the Zeeman effect is negligible (see SI), indicates the exponent $n$ in $\Delta\varrho(T) \sim T^n$. In addition to the AF phase boundary $B_N(p)$ at fields below 3 T, there are three metamagnetic phase boundaries at higher fields. All magnetic phase boundaries seem to merge at a potential multi-critical point near 0.8 GPa and 3 T. The AF existence range is embraced by the "Mott line", i.e., a line of true quantum phase transitions up to slightly below 1 GPa (solid), but a line of quantum crossovers at higher pressures (dashed); it separates the itinerant, heavy Fermi-liquid, regime in the upper, right part of the phase diagram from the local-moment paramagnetic and AF regimes in the lower left part. **(b)** Global phase diagram in the parameter space of quantum fluctuations of the local-moment magnetism ($G$) and the ratio of the Kondo to RKKY couplings ($J_K/I$). The shaded region is proposed to contain the qualitative projection of the combined $p - B$ tunings of (**a**) onto the $G - J_K/I$ plane (see text). Note that the parameter $G$, like $J_K$ and $I$, is specified at the Hamiltonian level and therefore well-defined for a Kondo-lattice model, regardless of whether the ground state is in the Kondo-screened or local-moment regime. Here the paramagnetic phases whose Fermi surface is large (from Kondo entanglement) and small (due to Kondo destruction) are denoted by $P_L$ and $P_S$, respectively, and their AF counterparts are labelled by $AF_L$ and $AF_S$.